\documentclass[11pt]{article}
\usepackage{moriond}
\usepackage{epsfig}

\def\Journal#1#2#3#4{{#1} {\bf #2}, #3 (#4)}

\def\PLB{{\em Phys. Lett.}  B}
\def\PRL{\em Phys. Rev. Lett.}
\def\PRD{{\em Phys. Rev.} D}

\def\be{\begin{equation}}
\def\ee{\end{equation}}
\def\bea{\begin{eqnarray}}
\def\eea{\end{eqnarray}}

\begin{document}
\vspace*{4cm}
\title{B physics prospects at LHCb}

\author{ T. LA\v{S}TOVI\v{C}KA }

\address{CERN, CH-1211 Gen\`{e}ve 23, Suisse/Switzerland}

\maketitle\abstracts{
The LHCb experiment at LHC is a single-arm spectrometer designed to pursue an extensive study of B physics and CP violation. In this contribution the physics which will be performed by LHCb is reviewed.}

\section{Introduction}

In last years B-factories did an outstanding job in delivering precise data to constrain the unitarity triangle within the Standard Model. However, there is room for improvements, such as the measurement of \mbox{$\gamma$ angle} or $B^0_s \bar{B}^0_s$ mixing, and new physics can still be hidden in the CKM picture.

At LHCb new physics could be observed indirectly in B decays, especially those involving box and penguin diagrams. Indirect approach can access high energy scales sooner and thus to see possible new physics effects earlier and, in principle, also the phases of the new couplings can be accessed. If new physics will be found at LHC in direct searches, B physics measurements will help to investigate the corresponding flavour structure.

Large Hadron Collider (LHC) at CERN is foreseen to deliver first pp collisions in year 2007. Compared to B-factories B physics at LHC has the great advantage of a high $b\bar{b}$ cross section ($\sigma_{bb} \sim 500\,\mu{\rm b}$) and of the production of all $b$-hadron species, including $B_s$, $b$-baryons and $B_c$. Another advantage is the possibility to reconstruct precisely the primary vertex as well as decay vertices due to a large number of tracks and a large $b$-hadron boost, respectively. On the other hand, high background (the inelastic cross section is $\sim 80\,{\rm mb}$), presence of underlying events and high particle multiplicity are among challenges. These features demand an excellent trigger capability, with good efficiency also on fully hadronic decay modes of $b$-hadrons, excellent tracking and vertexing performance, allowing for high mass and proper time resolutions, and excellent particle identification to separate exclusive decays.

The LHCb experiment is designed to maximise the B acceptance within given cost and space constraints. It is a forward spectrometer relying on relatively soft high $p_T$ triggers. Acceptance, given in terms of pseudo-rapidity, corresponds to \mbox{$1.9 < \eta < 4.9$}. It suits well for b events due to the forward peaked production of $b$-hadrons at LHC. One year of data-taking at a nominal luminosity of $2.10^{32}\,{\rm cm}^{-2}{\rm s}^{-1}$ corresponds approximately to annual integrated luminosity of $2\,{\rm fb}^{-1}$ and to production of $10^{12}$ $b\bar{b}$ events. The nominal luminosity is lower than that of ATLAS and CMS experiments since LHC beams are locally defocused in order to limit event pile-up. A detailed description of the LHCb detector can be found in reference [\,\cite{lhcb_detector}].

\section{B physics measurements at LHCb}

\subsection{Measurement of $sin 2\beta$}

Measurement of  $\sin 2\beta$ with the `golden' channel $B^0 \rightarrow J/\psi K_S$ is one of the first CP violation measurements foreseen at LHCb. It is not the main physics goal since it will be already very well measured by B-factories. However, measurement of $\sin 2\beta$ will be an important check of CP analyses and of tagging performance. Furthermore, it can search for direct CP violating term $\propto \cos \Delta m_dt$. Expected number of reconstructed $B^0 \rightarrow J/\psi K_S$ events is about 240.000 per accumulated luminosity of $2\,{\rm fb}^{-1}$ leading to a corresponding precision of $\sigma_{stat}(\sin 2\beta) \sim 0.02$ in one year of data taking. Currently $\sin 2\beta$ is known to a precision of $\sim 0.04$.

\subsection{Measurements of $B_s^0 \bar{B}_s^0$ mixing}

\subsubsection{$\Delta m_s$ measurement}

According to the existing knowledge, the $B^0_s \bar{B}^0_s$ oscillation was too fast to be resolved at LEP and SLC experiments. The Tevatron is at present the only available source of $B_s^0$ mesons and thus CDF and D0 have a chance to find a mixing signal in the near future. The measurement requires excellent performances in the event reconstruction and signal purity, proper time resolution and flavour tagging.

\begin{figure}
\centering\epsfig{file=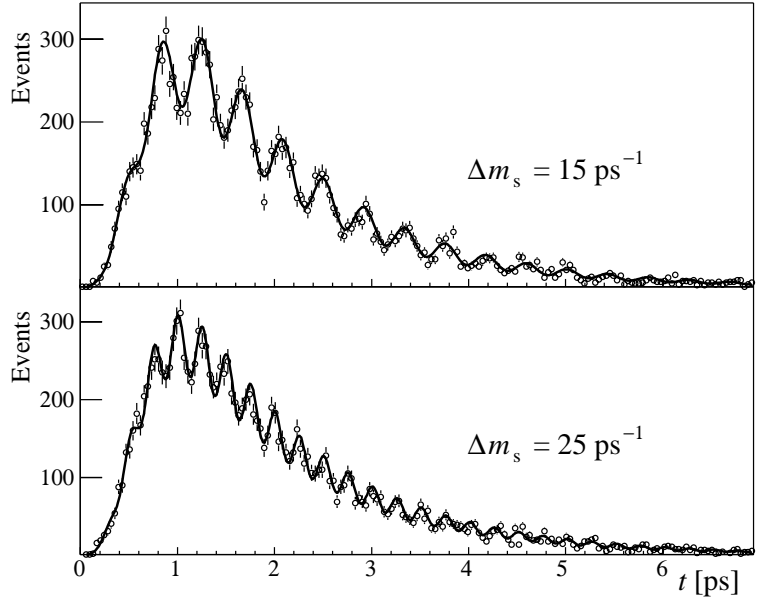, width=10cm}
\vspace{-0.35cm}
\caption{Simulated proper time distribution of $B_s^0 \rightarrow D_s\pi$ candidates which have been flavour tagged as having not oscillated. Plots correspond to one year of data-taking and two different values of $\Delta m_s$.}
\label{mixing_plot}
\end{figure}

A precise measurement of the $B^0_s \bar{B}^0_s$ mixing may come from the LHCb experiment. The best channel for these studies is $B^0_s \rightarrow D_s^-\pi^+$. Results of LHCb full simulation indicate a proper time resolution of $\sigma_\tau \simeq 40\,{\rm fs}$ and yield of 80.000 events per $2\,{\rm fb}^{-1}$ with a signal to background ratio of about 3. The effective efficiency for flavour tagging is estimated to be about 6\%. The simulated proper time distribution of tagged events is shown in Figure~\ref{mixing_plot} for two different values of $\Delta m_s$. In one year of data-taking a $5\sigma$ observation of oscillation is expected up to about $68\,{\rm ps}^{-1}$. Once observed, the estimated statistical precision to measure $\Delta m_s$ is $ \sim 0.01\,{\rm ps}^{-1}$ with a luminosity of $2\,{\rm fb}^{-1}$.

\subsubsection{$\phi_s$ and $\Delta\Gamma_s$ from $B_s \rightarrow J/\psi\phi$}

In the Standard Model the phase $\phi_s$ of the $B^0_s \bar{B}^0_s$ mixing is expected to be very small ($\phi_s = -\arg(V_{ts}^2) = -2\lambda\eta^2 \sim -0.04$) and thus sensitive to new physics in $b \rightarrow s$ transitions. Hints of new physics could be potentially seen also in the measurement of the decay width difference between the two CP eigenstates: \mbox{$\Delta\Gamma = \Gamma(B_L) - \Gamma(B_H)$}, which is expected to be about 10\% in the Standard Model. Both $\phi_s$ and $\Delta\Gamma_s$ can be measured from $B_s \rightarrow J/\psi(\mu\mu)\phi(KK)$ decays. For $2\,{\rm fb}^{-1}$ of accumulated data LHCb expects to collect about 125.000 $J/\psi(\mu\mu)\phi$ events which would allow to measure $\sin(\phi_s)$ and $\Delta\Gamma_s/\Gamma_s$ with statistical precision of $0.031$ and $0.011$, respectively. After first five years of running and addition of pure CP channels the sensitivity to $\sin(\phi_s)$ could reach $0.013$.

\subsection{Prospects on $\gamma$ measurements}

\subsubsection{$B_s \rightarrow D_sK$ decays}

\begin{figure}
\raisebox{2cm}{\epsfig{file=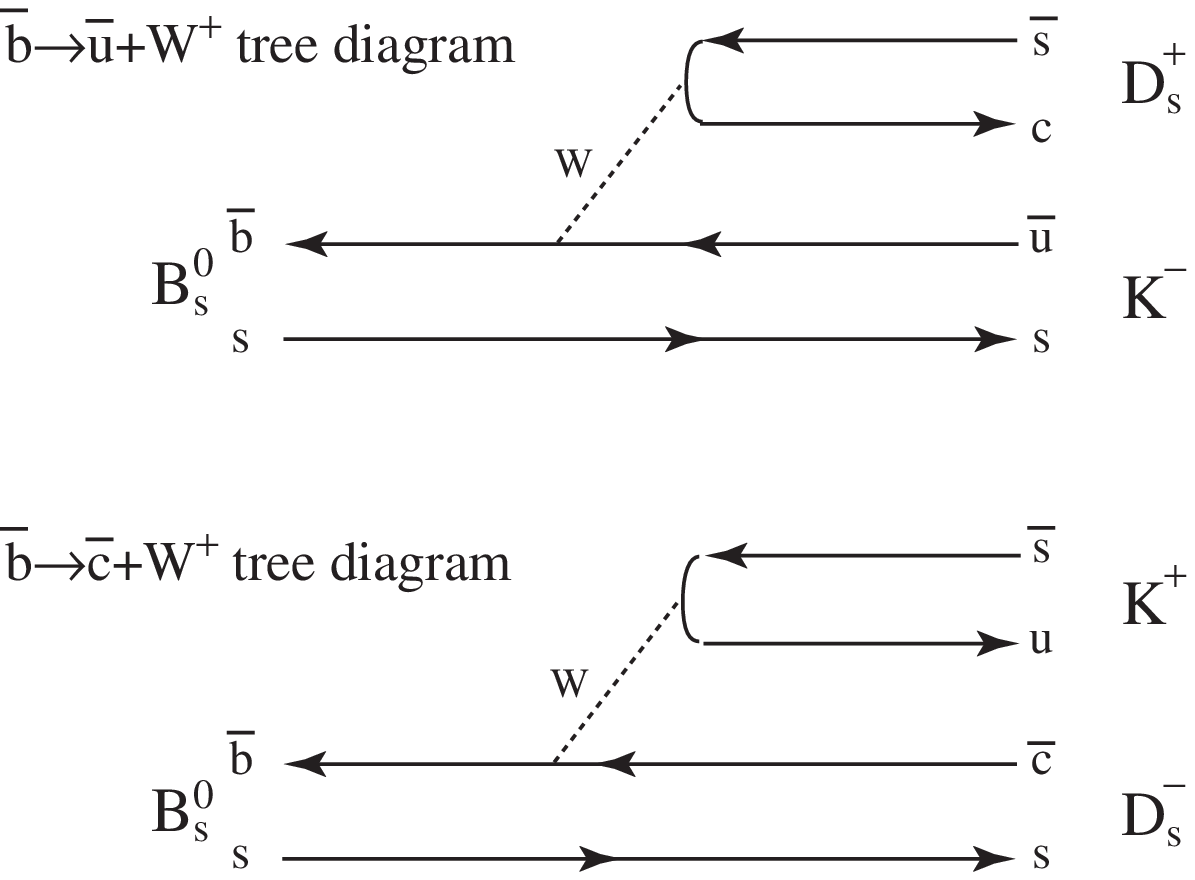, width=6.5cm}}
\epsfig{file=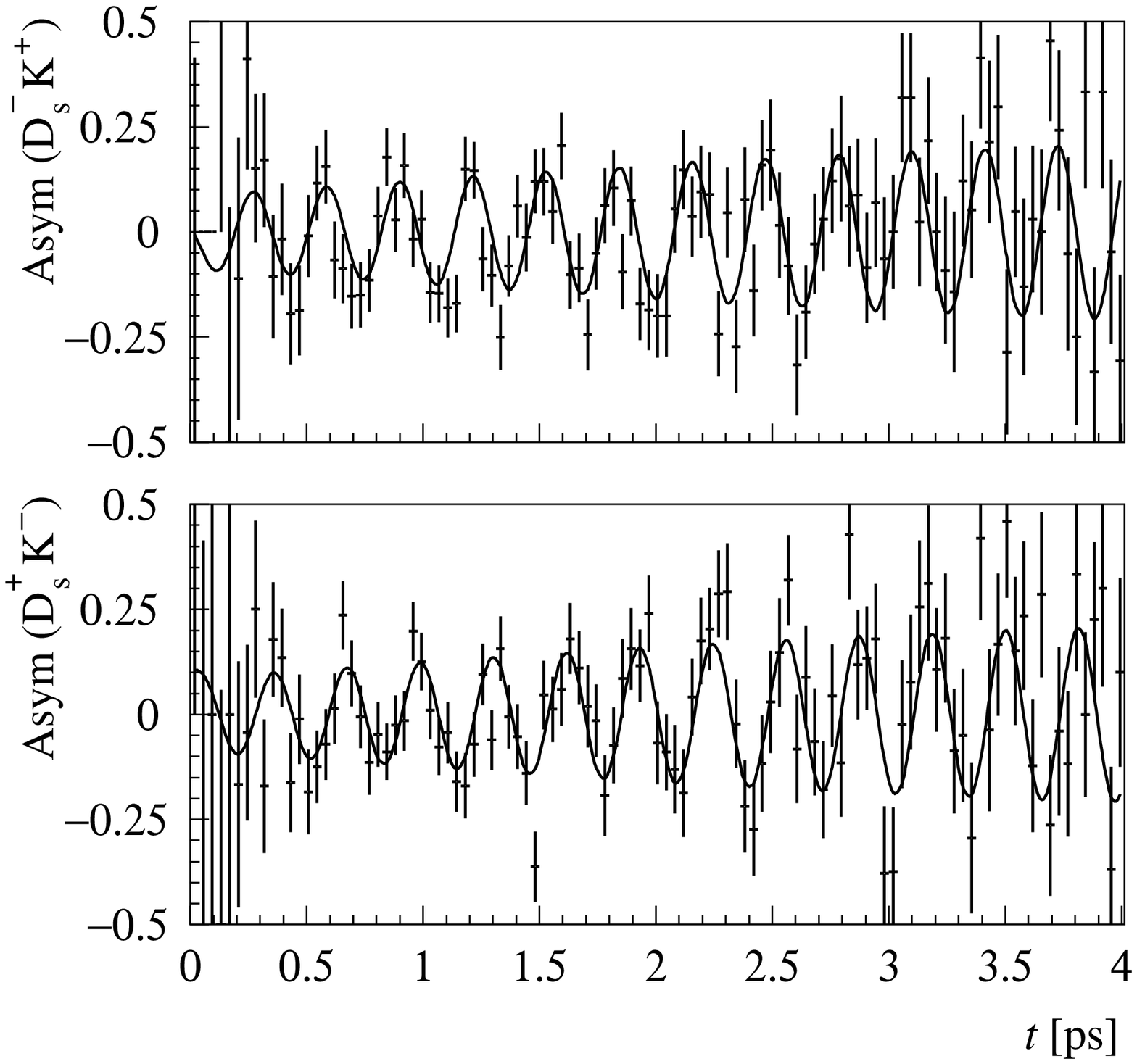, width=8.5cm}
\vspace{-0.35cm}
\caption{Feynman tree diagrams of $B_s^0 \rightarrow D_s^+K^-$ and $B_s^0 \rightarrow D_s^-K^+$ (left). Asymmetry of $B^0_S\bar{B}^0_s$ as a function of proper time (right) for $D_s^-K^+$ (top) and $D_s^+K^-$ (bottom) candidates. Simulations correspond to $\Delta m_s = 20\,{\rm ps}^{-1}$ and five years of data-taking.}
\label{gamma_plot}
\end{figure}

Tree diagrams corresponding to $B_s^0 \rightarrow D_s^+K^-$ and $B_s^0 \rightarrow D_s^-K^+$ are shown in Figure~\ref{gamma_plot} (left). The two decay channels can interfere via mixing. The phase $\gamma+\phi_s$ and the strong phase $\Delta$ can be extracted from the measurement of the time-dependent decay asymmetries, which is illustrated in Figure~\ref{gamma_plot} (right). Furthermore, if $\phi_s$ is determined from another measurement, for instance from $B_s \rightarrow J/\psi\phi$, the angle $\gamma$ can be extracted with a very small theoretical uncertainty to a precision of $\sigma(\gamma) \sim 14^\circ$ in one year of data-taking, assuming $\Delta m_s = 20\,{\rm ps}^{-1}$. Decays are not sensitive to new physics and thus the measured $\gamma$ corresponds to the Standard Model $\gamma$.

\subsubsection{$B^0 \rightarrow D^{0}K^{\star 0}$ decays}
 
A theoretically clean way to measure $\gamma$ is $B^0 \rightarrow D^{0}K^{\star 0}$. This decays are represented by tree diagrams and thus, as $B_s \rightarrow D_sK$, have no sensitivity to new physics. The method [\,\cite{d0ks0}] is based on a measurement of six time integrated decay rates for $B^0_d \rightarrow D^{0}K^{\star 0}, \bar{D}^{0}K^{\star 0}, D^{0}_{CP}K^{\star 0}$ and their CP conjugates. Signal amounts to about 4.000 events per $2\,{\rm fb}^{-1}$ of data and corresponding precision of measured $\gamma$ is $\sim 8^\circ$. In this measurement also the strong phase $\Delta$ can be accessed.

\subsubsection{$B^\pm \rightarrow DK^\pm$ decays}

A newly proposed clean measurement of $\gamma$ involves $B^\pm \rightarrow DK^\pm$ tree decays. It is based on ADS method [\,\cite{ads}]. The idea is to measure relative rates of $B^+ \rightarrow DK^+$ and $B^- \rightarrow DK^-$ decays with neutral $D$'s in the final state. The method is a candidate for statistically most precise determination of $\gamma$ at LHCb. The expected precision is about $\sigma_{stat}(\gamma) \sim 5^\circ$ after one year of data-taking ($2\,{\rm fb}^{-1}$). 

\subsubsection{$B^0_d \rightarrow \pi^{+}\pi^{-}$ and $B^0_s \rightarrow K^{+}K^{-}$}

Another set of channels of interest [\,\cite{ppkk,ppkk2}] for the study of $\gamma$ is the two-body decays $B^0_d \rightarrow \pi^{+}\pi^{-}$ and $B^0_s \rightarrow K^{+}K^{-}$. The decays involve loops, in the penguin diagrams, thus the measurement of $\gamma$ is sensitive to new physics that might appear in the loops. A comparison to the values of $\gamma$ extracted with new physics insensitive methods will allow to pinpoint any new physics contribution. Expected yields are 26.000 and 37.000 events per accumulated luminosity of $2\,{\rm fb}^{-1}$ for $B^0_d \rightarrow \pi^{+}\pi^{-}$ and $B^0_s \rightarrow K^{+}K^{-}$, respectively. This would lead to a precision of $\sigma(\gamma) \sim 5^\circ$.

\subsection{Rare decays: $B_s \rightarrow \mu^+\mu^-$}

This rare decay involves flavour changing neutral currents whose branching ratio is very low and estimated to be $BR(B_s \rightarrow \mu^+\mu^-) = (3.5\pm 0.1)\times 10^{-9}$ in the Standard Model [\,\cite{mumu}]. In various
SUSY extensions of the Standard Model it can be enhanced by roughly two orders of magnitude. The best upper limit on the branching ratio at present come from experiments at Tevatron and reaches few$\,\times\,10^{-7}$ at 95\% CL. The LHCb experiment has prospects for a significant measurement and aims at a $2\sigma$ measurement in 2 years.

\section{Conclusion}

In the coming years CP asymmetries will be measured at LHCb using several $B^0_d$ and $B^0_s$ mesons and $b$-baryons decay channels. Very rare decays will also be studied, thanks to the high $b\bar{b}$ cross section available. This program is complementary to B-factories and will allow to complete and improve the available results and possibly to reveal first signals of new physics.

\end{document}